\documentclass[aps,10pt,prd,twocolumn,preprintnumbers,amsmath,amssymb,nofootinbib,superscriptaddress,a4paper,showpacs]{revtex4-1}

\usepackage[breaklinks]{hyperref}
\usepackage[latin1]{inputenc}
\usepackage{graphicx}
\usepackage{color}
\usepackage{amsfonts,amsthm}
\usepackage{bm,bbm}

\newcommand{\be}{\begin{equation}}
\newcommand{\ee}{\end{equation}}
\newcommand{\ba}{\begin{eqnarray}}
\newcommand{\ea}{\end{eqnarray}}
\def\bs{\begin{subequations}}
\def\es{\end{subequations}}
\def\a{\alpha}
\def\b{\beta}
\def\de{\delta}

\def\Om{\Omega}
\def\om{\omega}

\def\s{\sigma}

\def\N{\nabla}
\def\cA{\mathcal{A}}

\def\cL{\mathcal{L}}

\def\cP{\mathcal{P}}

\def\cS{\mathcal{S}}

\def\cV{\mathcal{V}}

\def\ds{d_{\rm S}}
\def\dh{d_{\rm H}}
\def\dw{d_{\rm W}}

\def\p{\partial}

\def\B{\Box}
\newcommand{\Eq}[1]{(\ref{#1})}
\def\com{\color{magenta}}
\def\cob{\color{blue}}
\newcommand{\oarX}[1]{\href{http://arxiv.org/abs/#1}{{\ttfamily\com arXiv:#1}}}
\newcommand{\arX}[1]{\href{http://arxiv.org/abs/#1}{{\ttfamily\com arXiv:#1}}}
\newcommand{\doin}[6]{\href{http://dx.doi.org/#1}{{\cob #2 #3 {\bf #4}, #5 (#6)}}}
\newcommand{\doinn}[5]{\href{http://dx.doi.org/#1}{{\cob #2 {\bf #3}, #4 (#5)}}}
\newcommand{\doij}[5]{\href{http://dx.doi.org/#1}{{\cob #2 #3 (#5) #4}}}

\newcommand{\ndoinn}[5]{\href{#1}{{\cob #2 {\bf #3}, #4 (#5)}}}
\newcommand{\tia}[1]{}

\def\lp{\ell_{\rm Pl}}
\def\ep{E_{\rm Pl}}
\def\rme{e}
\def\rmd{d}
\def\rmi{i}

\begin{document}

\title{Black-hole entropy and minimal diffusion}

\author{Michele Arzano}
\affiliation{Dipartimento di Fisica and INFN, ``Sapienza'' University of Rome, P.le A. Moro 2, 00185 Roma, Italy}

\author{Gianluca Calcagni}
\affiliation{Instituto de Estructura de la Materia, CSIC, Serrano 121, 28006 Madrid, Spain}

\date{July 23, 2013}

\begin{abstract}
The density of states reproducing the Bekenstein-Hawking entropy-area scaling can be modeled via a nonlocal field theory. We define a diffusion process based on the kinematics of this theory and find a spectral dimension whose flow exhibits surprising properties. While it asymptotes four from above in the infrared, in the ultraviolet the spectral dimension diverges at a finite (Planckian) value of the diffusion length, signaling a breakdown of the notion of diffusion on a continuum spacetime below that scale. We comment on the implications of this minimal diffusion scale for the entropy bound in a holographic and field-theoretic context.
\end{abstract}




\preprint{\doin{10.1103/PhysRevD.88.084017}{Phys.\ Rev.}{D}{88}{084017}{2013} \hspace{10.5cm} \arX{1307.6122}}

\maketitle

\section{Introduction} 

Evidence has been accumulating in recent years suggesting that at small scales the dimensionality of spacetime might flow to values different than four due to quantum effects of the geometry \cite{tHo93}. The notion of \emph{spectral dimension} can be used to explore spacetime geometries beyond the usual picture of smooth manifolds which can emerge in quantum gravity scenarios. Its running to lower values in the ultraviolet is a common feature of several approaches \cite{flow,LaR5,nonc,Hor3,Mod11,frc7,CES}. In general, whenever a suitable generalization of the Laplacian operator is available one can consider a diffusion process via a heat equation and the return probability allows one the definition of the spectral dimension. This probe of nonconventional geometries turned out to be useful also in the context of various quantum gravity--inspired field-theoretic models with Lorentz symmetry breaking, higher-order or modified derivatives, or nonlocal Lagrangians (e.g., \cite{Mod11,frc7,frc4}).

In this paper, we explore the diffusion properties of a nonlocal field theory which effectively models the degrees of freedom of the quantum geometry of a black hole. In \cite{Pad98,Pad99}, it was argued that black holes provide a general setup to probe the quantum microstructure of spacetime. In this scenario, the entropy-area law of black holes \cite{BH} originates from quantum-gravitational degrees of freedom. The entropy of a Schwarzschild black hole of mass (mean energy) $M=E\gg \ep$, proportional to the area $A$ of the event horizon, can be reproduced to leading order by a density of energy states
\be\label{dos}
\rho(E)=\exp[4\pi (E/\ep)^2]+O(E/\ep)\,,
\ee
where $\ep$ is Planck's energy. In fact, $S:=\ln\rho(E)=A/(4\lp^2)$, where $\lp=\ep^{-1}$ is the Planck length. This density of states can be reproduced by a very generic effective field-theory model $\cS=\int\rmd t\,\rmd^{D-1}{\bf x}\cL$ with a nonlocal self-interaction in space:
\be
\cL=\frac12\dot\phi^2(t,{\bf x})-\frac12\int\rmd^{D-1}{\bf y}\,\phi(t,{\bf x})\,\Om^2({\bf y}-{\bf x})\,\phi(t,{\bf y})\,,\label{es}
\ee
where $D$ is the number of topological dimensions (for concreteness, we can take $D=4$), $\phi$ is some effective field encoding the microscopic degrees of freedom of the quantum geometry, and the form factor $\Om^2$ acts as a smearing of the fields over a spatial distance $\sim\lp$. Equation \Eq{es} is assumed to be valid at microscopic scales not much larger than $\lp$. Apart from this ansatz, the fundamental degrees of freedom and the details of the underlying theory of quantum gravity are otherwise unspecified. 

In order to study a diffusion process governed by the kinematics of the nonlocal action above, we start by noticing that in momentum space the Fourier transform of the form factor $\Om^2$ yields a modified dispersion relation
\be\label{dire0}
\tilde F(k^0,{\bf k}):= -k_0^2+\om^2({\bf k})=0\,.
\ee
An example of a Lorentz-breaking dispersion relation leading to Eq.\ \Eq{dos} is
\be
\om^2({\bf k}):=\frac{(D-1)\ep^2}{8\pi}\,\ln\left[1+\frac{8\pi |{\bf k}|^2}{(D-1)\ep^2}\right]\,,\label{dire}
\ee
as one can check \cite{Pad98,Pad99} from the definition of the thermodynamical partition function $Z(\b)\propto\int \rmd{\bf k}\,\rme^{-\b \om({\bf k})}=\int\rmd E\,\rho(E)\,\rme^{-\b E}$, where $\b$ is the inverse temperature. The form factor $\Om^2$ in position space can be found by antitransforming Eq.\ \Eq{dire}. For instance, in $D=2$ one gets $\Om^2(y-x)=-(2\s_*|y-x|)^{-1}\exp(-|y-x|/\sqrt{2\s_*})$ \cite{Pad98},
where we introduced the critical squared length (the area of a disk of Planck radius)
\be\label{crit}
\s_*=4\pi\lp^2\,.
\ee
Thus, fields are smeared over a region of size $\sqrt{2\s_*}=O(\lp)$. In $D$ dimensions, $\Om^2\propto -(r_*/r)^{(D-1)/2}K_{(D-1)/2}(r/r_*)$, where $r=|{\bf x}-{\bf y}|$ and $K$ is the modified Bessel function of the second kind \cite{Pad99}. Again, the effective correlation length is $r_*:= \sqrt{2\s_*/(D-1)}=O(\lp)$.

The variables $t$ and ${\bf x}$ in Eq.\ \Eq{es} {do not} have to be the spacetime coordinates: they can also represent variables in the abstract space of the putative microscopic quantum spacetime theory \cite{Pad98}. 
In this paper, however, we take Eq.\ \Eq{es} at face value and study in more detail the claim that scales below the Planck length cannot be probed in such an effective theory. In principle, the smearing via the form factor $\Om^2$ could also lead to a sort of Planck-scale ``fuzziness'' of spacetime, where geometry can still be described by conventional indicators such as the spectral dimension. In other words, it is not obvious whether the correlation length $\sim\sqrt{\s_*}=O(\lp)$ acts as a watershed between two spacetime regimes or as a lower bound for length measurements. Here, we will show that the model is constructed in such a way that there is no manifold structure below the Planck scale. The Planck length is a minimal physical scale to all purposes, and it becomes meaningless to ask how spacetime is modified at smaller distances.

Let us stress that there are only two key requirements beyond our main result, both stated in \cite{Pad98}. The first is the presence of a hypersurface with infinite redshift. Roughly speaking, the event horizon ``stretches'' virtual high-energy field excitations (representing the various interactions of matter with quantum geometry) to sub-Planckian energies and allows them to become real modes, which then populate thermodynamical energy levels. The second is Eq.\ \Eq{dos}. \emph{Any} microscopic theory of quantum gravity with the correct effective density of states \Eq{dos} for a black hole, i.e., predicting the entropy-area law (string theory \cite{strbh} and loop quantum gravity \cite{lqgbh} are examples), will be described (after some coarse-graining approximation) by a nonlocal effective model of the above or similar form near the horizon, with correlation functions displaying a universal short-scale behavior. As argued in \cite{Pad98}, it is not necessary to know the details of the ultimate theory, assuming it exists, to reproduce some basic thermodynamical properties. In this sense, our conclusions on the spectral dimension will not hold in scenarios violating our working hypotheses, but otherwise they will be quite general.

\section{Nonlocality, diffusion, and spectral dimension}
 
To probe the local structure of spacetime, we Euclideanize coordinate time $t$, $t\to-\rmi t= x_D$, and let a pointwise test particle diffuse starting from some spacetime point $x'=(x_D',{\bf x}')$. We then ask what is the probability $P$ to find the particle at another point $x=(x_D,{\bf x})$ after some abstract diffusion ``time'' $\s$ has elapsed. This process is encoded in a diffusion equation. For instance, the ordinary diffusion equation for Minkowski spacetime reads $(\p_\s-\B_{\rm E})P=0$, where $\B_{\rm E}=\p_D^2+\N^2$ and $\N^2$ is the spatial Laplacian; this yields an ordinary Brownian motion with Gaussian probability density function $P$ and, eventually, a spectral dimension $\ds=D$.

In our case, we have to replace the standard Laplacian with a nonlocal derivative operator, reproducing, in momentum space, the dispersion relation \Eq{dire0} with Eq.\ \Eq{dire}. It is easy to convince oneself that the diffusion equation should be of the form
\be\label{die}
\left[\p_\s-F(\rmi\p_D,\N)\right]P(x,x',\s)=0\,,
\ee
where
\be\label{efen}
F(\rmi\p_D,\N)=\frac{\p^2}{\p x_D^2}-\frac{D-1}{2\s_*}\,\ln\left[1-\frac{2\s_*}{D-1}\,\N^2\right].
\ee
To see this, we notice that a nonlocal interaction can always be expressed as a nonlocal kinetic term \cite{BOR}. In fact (time dependence in $\phi$ omitted), $\int\rmd{\bf y}\,\Om^2({\bf y}-{\bf x})\,\phi({\bf y})=\int\rmd{\bf z}\,\Om^2({\bf z})\,\phi({\bf z}+{\bf x})=[\int\rmd{\bf z}\,\Om^2({\bf z})\,\rme^{{\bf z}\cdot \N_x}]\phi({\bf x})$. 
Taking the Fourier transform of $\rme^{{\bf z}\cdot \N_x}$ and using the dispersion relation \Eq{dire0}, we get
\ba
\int\rmd{\bf z}\,\Om^2({\bf z})\,\rme^{{\bf z}\cdot \N_x}&=&\int\rmd{\bf z}\,\Om^2({\bf z})\,\int\rmd{\bf k}\,\rme^{\rmi {\bf k}\cdot {\bf z}}\de({\bf k}-\rmi\N_x)\nonumber\\
&=&\int\rmd{\bf k}\left[\int\rmd{\bf z}\,\Om^2({\bf z})\,\rme^{\rmi {\bf k}\cdot {\bf z}}\right]\de({\bf k}-\rmi\N_x)\nonumber\\
&=&\int\rmd{\bf k}\,\om^2({\bf k})\,\de({\bf k}-\rmi\N_x) \nonumber\\
&=& \om^2(\rmi\N_x)\,,
\ea
which yields \Eq{efen} when adopting Eq.\ \Eq{dire}. The operator \Eq{efen} is mathematically well defined \cite{NY} and admits a series representation with finite coefficients [in general, even well-defined nonlocal operators do not, e.g., $(\N^2)^\a$ with $\a$ complex]:
\be\label{seri}
F(\rmi\p_D,\N)=\frac{\p^2}{\p x_D^2}+\frac{D-1}{2\s_*}\sum_{m=1}^{+\infty}\,\frac1m\left(\frac{2\s_*}{D-1}\,\N^2\right)^m.
\ee
In the limit $\s_*\to 0$ ($\lp\to 0$, large diffusion scales), $F\to\B_{\rm E}$, as one can see from Eq.\ \Eq{seri}. 

The solution of Eq.\ \Eq{die} is
\be\label{dies}
P(x,x',\s)=\int \frac{\rmd^D k}{(2\pi)^D}\,\rme^{-\s \tilde F(\rmi k^0,{\bf k})}\,\rme^{\rmi k\cdot (x-x')}\,.
\ee
This expression can be computed exactly. The integral in $k^0$ yields the usual Gaussian normalization $1/\sqrt{4\pi\s}$, while the integral in spatial momenta can be done in polar coordinates. The result is
\ba
P(x,x',\s) &=& \frac{\rme^{-\frac{|x_D-x_D'|^2}{4\s}}}{\sqrt{4\pi\s}}\frac{2^{1-\frac{D-1}{2}\frac{\s}{\s_*}}}{\Gamma\left(\frac{D-1}{2}\frac{\s}{\s_*}\right)}\left(\frac{D-1}{4\pi\s_*}\right)^{\frac{D-1}{2}}\label{P}\\
&&\times\left(\frac{r}{r_*}\right)^{\frac{D-1}{2}\left(\frac{\s}{\s_*}-1\right)} K_{\frac{D-1}{2}\left(\frac{\s}{\s_*}-1\right)}\left(\frac{r}{r_*}\right)\,, \nonumber
\ea
where $\Gamma$ is Euler's function and $\s_*$ is given by Eq.\ \Eq{crit}. Instead of imposing an initial condition for \Eq{dies}, we fixed the normalization of $\int\rmd^Dx\,P$ to 1. At $\s=0$ we do not get the usual delta, due to the smearing effect: $P(x,x',\s\sim 0)\sim-\de(x_D-x_D')\s\Om^2(r)\neq\de(x-x')$. From Eq.\ \Eq{P}, we can extract much information. First, $P$ defines probabilistic expectation values of the form $\langle f(x)\rangle=\int_{-\infty}^{+\infty}\rmd^Dx\, P(x,0,\s)\,f(x)$, with $\langle 1\rangle=1$. In particular, the mean squared displacement is 
\be\label{msd}
\langle x^2\rangle=\langle x_D^2+r^2\rangle=2D\s\,, \qquad x\neq 0\,,
\ee
and diffusion is nonanomalous (the walk dimension determined by $\langle x^2\rangle\propto \s^{2/\dw}$ is equal to 2). However, it is not ordinary, either. The trace of $P$ in position space is the return probability $\cP(\s):=\int\rmd^Dx\,P(x,x,\s)$.
At small $z$, $z^\nu K_{\pm\nu}(z)\sim 2^{\nu-1}\Gamma(\nu)$ if $\nu\neq 0$, so that
\be\label{repr}
\cP(\s)=\cA\,\sqrt{\frac{\s_*}{\s}}\,\frac{\Gamma\left[\frac{D-1}{2}\left(\frac{\s}{\s_*}-1\right)\right]}{\Gamma\left(\frac{D-1}{2}\frac{\s}{\s_*}\right)}\,,\qquad \s> \s_*\,,
\ee
where $\cA$ is a divergent constant proportional to the total volume $\cV$. This expression diverges at $\s=(1-2n)\s_*$, where $n\in\mathbb{N}$. Since $\s>0$, the only singular point of interest is $\s=\s_*$ ($n=0$). At scales $\s<\s_*$, the return probability is no longer positive semidefinite, implying that $P$ is not a probability density function at coincident points $x=x'$. Taking $x\sim x'$ means probing infinitely close points within an infinitely small diffusion distance $\delta\sqrt{\sigma}$; but at scales $\s<\s_*$, the diffusion process is ill defined. Consequently, in this range of scales \emph{there is no diffusive process by which the spectral dimension could be defined}. As $\cP$ should also be continuous, it is natural to interrupt the process at the largest pole of Eq.\ \Eq{repr}, i.e., at $\s=\s_*$. It is here where the correct initial condition $P(x,x',\s_*)$ for the stochastic process is set \emph{a posteriori}. Positivity of the solution of the diffusion equation at all initial points is a strong criterion characterizing an effective quantum geometry, which not only consolidates the determination of the number $\ds$ on physical grounds, but also constitutes a finer tool to classify geometries with the same spectral dimension \cite{frc4,frc7,CES}.

The presence of a minimal diffusion length $\ell_*=\sqrt{\s_*}$ suggests that physical happenings cannot be separated by time-space scales smaller than $\ell_*$, and that we are actually facing a discreteness effect. Spacetime near a black hole shows an effective discrete structure at microscopic scales. It is exciting to notice that this picture is compatible with the holographic principle: a discrete structure depletes spacetime (and the phase space of the system) of degrees of freedom which would otherwise contribute to the black-hole entropy proportionally with the volume.

From Eq.\ \Eq{repr}, we get the analytic expression for the spectral dimension:
\ba
\ds(\s)&:=&-2\,\frac{\p\ln\cP(\s)}{\p\ln\s}\nonumber\\
&=&1+(D-1)\frac{\s}{\s_*}\left\{\psi\left(\frac{D-1}{2}\frac{\s}{\s_*}\right)\right.\nonumber\\
&&\left.-\psi\left[\frac{D-1}{2}\left(\frac{\s}{\s_*}-1\right)\right]\right\}\,,\quad\, \s> \s_*\,,\label{ds}
\ea
where $\psi(a)=\p_a\Gamma(a)/\Gamma(a)$ is the digamma function.

\begin{figure}
\centering
\includegraphics[width=7.5cm]{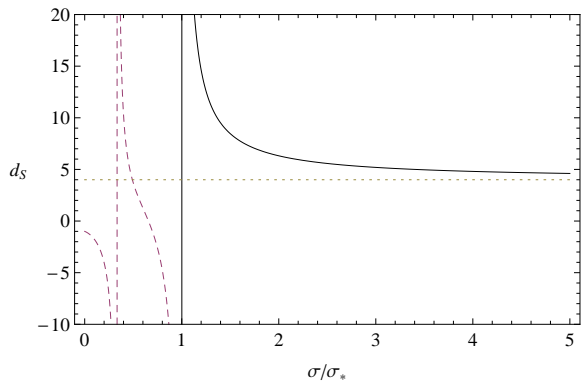}
\caption{\label{fig1} Spectral dimension for $D=4$. In the forbidden region $\s<\s_*$, the function \Eq{ds} (dashed curve) acquires an increasing number of poles as $D$ increases. This function, however, cannot be regarded as the spectral dimension.}
\end{figure}

Figure \ref{fig1} shows the whole profile \Eq{ds} in $D=4$ for $\s>0$, with the understanding that the dashed part for $\s<\s_*$ is reported only for illustrative purposes. Asymptotically,
\be\label{dsasi}
\ds\sim \begin{cases} +\infty &(\s=\s_*^+)\\
D &(\s\gg\s_*)\end{cases}\,.
\ee
While in the infrared $\ds$ tends to the topological dimension, at the critical minimal length scale $\ell_*$ it diverges. This confirms quantitatively the limitation in measuring times and lengths incorporated in the framework of \cite{Pad98,Pad99}. Notice that in Eq.\ \Eq{die} we chose the diffusion coefficient $\ell$ (a length) in front of $F$ to be equal to 1, so that $\s$ has dimension (length)$^2$. If we had defined units so that the critical scale $\s_*\to 4\pi\lp^2/\ell$ were a length, taking $\ell=O(1)\lp$ would have led to the same minimal length $\ell_*$ as above, modulo an immaterial $O(1)$ prefactor which can always be reabsorbed in the diffusion parameter $\s$.

An anomalous spectral dimension is compatible with the normal walk dimension $\dw=2$ obtained from Eq.\ \Eq{msd} because, contrary to fractals, $\dw\neq 2\dh/\ds$, where $\dh$ is the Hausdorff dimension (in this case, $\dh=D$). This is expected, since the density of states \Eq{dos} \cite{Pad98} is not the one met in fractals \cite{frc7}.

\section{Discussion} 

To the best of our knowledge, in the context of effective models of quantum gravity, this is the first example of a profile for the spectral dimension which stops at a minimal diffusion length. There are cases where geometry possesses some characteristic scale $\ell_{\rm crit}$, which, however, is not minimal. Models of ``fuzzy manifolds'' have a transition at a critical length $\ell_{\rm crit}$ where the Hausdorff dimension $\dh=2-(D-\ds)$ becomes negative \cite{NiN}, a feature which may have some connection with results in multifractal geometry \cite{Man}. However, the spectral dimension $\ds=\s D/(\s+\ell_{\rm crit}^2)$ falls to $\ds\sim 0^+$ all the way down to vanishing diffusion scale $\s\to 0$ \cite{MoN}. Also in asymptotic safety, the intrinsic fuzziness of the quantum geometry \cite{RSc} does not imply a minimal diffusion scale and the limit $\ds(\s\to 0)$ is well defined \cite{LaR5,CES}. Finally, in noncommutative spacetimes (where $\ell_{\rm crit}=\lp$), the spectral dimension changes with the scale but geometry can be probed to arbitrarily small lengths \cite{nonc}, and $\lp$ acts as a smearing length rather than a cutoff. The functional form of dispersion relations in noncommutative momentum spaces is similar to Eq.\ \Eq{dire}, but it does not have the same ultraviolet limit \cite{AM}.

Here, on the other hand, the Planck scale plays the role of a \emph{minimal} rather than characteristic scale. Below it, we do not have a gradual loss of resolution of the diffusing probe, as in \cite{MoN,RSc}: simply, there is no diffusion at all. Whether one interprets it either as the absence of a continuum spacetime below $\lp$ or as an operational limitation in measuring scales with accuracy greater than $\lp$, the net result for physical observations is the same. 

The loss of Lorentz invariance at microscopic scales [Eqs.\ \Eq{es} and \Eq{dire0}] is therefore expected in frameworks where $\lp$ acts as a minimal diffusion scale. Forfeiting special relativity at these scales and recovering it in the infrared is not a unique feature of this model, and can be found also in other continuum scenarios with dimensional flow (such as Ho\v{r}ava-Lifshitz gravity \cite{CES,Hor3} or multiscale spacetimes with modified derivatives \cite{frc7}). On the other hand, nonlocal Laplacians are known to lead, in general, to unconventional geometric structures below a certain scale \cite{nonc,Mod11,frc4}, but by itself nonlocality is not the cause of the unique behavior we found. Even small deviations from Eq.\ \Eq{dire} give rise to a change in the density of states and, hence, to a spoiling of the entropy-area law. Thus, Eq.\ \Eq{dire} is perhaps less a toy model than deemed in \cite{Pad98}.

The topic of black holes and dimensional flow has been previously considered in \cite{CaG,Mur12}. There is little intersection between those results and ours. In \cite{Mur12}, dimensional flow is simply used as a motivation to study lower-dimensional black holes. 
In \cite{CaG}, dimensional flow was \emph{assumed} to be monotonic from $\ds\sim 4$ in the infrared to some value smaller than 3 in the ultraviolet. In that case, it was argued that a Schwarzschild black hole, described in a \emph{local} theory, stops evaporating when its radius shrinks to a minimal scale at which $\ds\sim 3$, below which the properties of the black hole can no longer be probed. Also, no observer, inside or outside the black hole, can see a value $\ds<2$. Here, we did not postulate a profile for $\ds$, but we started with a nonlocal theory realizing the density of states necessary to obtain the black-hole entropy-area law. The resulting spectral dimension is always \emph{greater} than 4. We did find a minimal scale $\ell_*\sim \lp$ below which it is not possible to check the spectral dimension of spacetime, but scales $\s\sim \s_*$ correspond to a geometry with $\ds\to\infty$. 

The divergence of the spectral dimension is somewhat difficult to assess, as the heat kernel \Eq{repr} neither resembles the case of ordinary manifolds nor reproduces the results for fractals. Still, we can advance an explanation by recalling that the definition of $\ds$ expresses the heat kernel as an effective power law proportional to the volume $\cV$ of the system, $\cP(\s)\sim\cV \s^{-\ds(\s)/2}$. In the limit $\s\to\s_*$, $\cP\sim \cV \s_*^{-\ds/2}\to 0$ instead of diverging as usual. This is encouragingly compatible with the holographic principle: the volume of the system is no longer the leading contribution in the heat kernel. 

We conclude with the following observation. In asymptotically flat spacetimes stable against gravitational collapse, the entropy $S$ of the truncated Fock space of bosonic and fermionic
 \emph{local} field theories is bounded from above by $A^{3/4}$, where $A$ is the boundary area of the region where the quantum fields live \cite{Yur}. This reproduces the bound by 't Hooft \cite{tHo93}. A key assumption to obtain this result is that the energy of the Fock states does not exceed an upper limit, conventionally fixed to be the Planck scale: $E<\ep$. On the other hand, in the presence of a black hole the entropy follows the area law $S\propto A$ and, according to the model discussed here, the correct description is in terms of a \emph{nonlocal} field theory. Nonlocality implies that these fields do not represent particles: the propagator of the theory \Eq{es} is $1/\tilde F$, the off-shell inverse of the dispersion relation \Eq{dire0}, which has two branch cuts with branch points at $k^0=\pm\om({\bf k})$. (The presence of branch cuts in the propagator and the consequent loss of the particle interpretation often occur in nonlocal theories \cite{BOR,doA92}, but not always \cite{BKcu}.)
 Here we point out a suggestive way to show that the two scenarios are, in fact, compatible. If we start from the nonlocal theory and truncate the dispersion relation for small momenta $|{\bf k}|\ll \ep$, to leading approximation we get a local theory with $\om^2({\bf k})=|{\bf k}|^2+O(|{\bf k}|^4)$. Then, we can consider the Fock space of this effective field theory but with field modes with momenta no larger than the Planck energy. This is precisely the situation where one meets the requirements for the nonholographic bound $S\leq (A/\lp^2)^{3/4}$. Thus, we conjecture that the discrepancy between the $A$ and $A^{3/4}$ laws lies in the infinite number of degrees of freedom thrown away by the truncation of the nonlocal dispersion relation. How the quasiparticle states of the fully nonlocal effective theory contribute to the entropy of the system has been outlined already in \cite{Pad98,Pad99}. However, a nontrivial check of this conjecture would go beyond classical thermodynamical considerations and enter the realm of quantum field theory, linking with the results of \cite{Yur}. This study will entail the management of an infinite number of particle fields with techniques outside the scope of this paper.

\section*{Acknowledgments}

The work of M.A.\ is supported by the E.U.\ Marie Curie Actions through a Career Integration Grant and in part by the John Templeton Foundation. The work of G.C.\ is under a Ram\'on y Cajal contract.

\end{document}